\documentstyle[prl,aps,multicol,psfig,epsfig]{revtex}

\begin{document}
\draft

\author{V.\ M.\ Pudalov$^{a,b}$, M.\ E.\ Gershenson$^{a}$, H.\ Kojima$^{a}$,
G.\ Brunthaler$^c$, A.\ Prinz$^c$, and G.\ Bauer$^c$}
\address{$^{a}$ Serin Physics Lab, Rutgers University, Piscataway NJ-08854, USA \\
          $^{b}$ P.\ N.\ Lebedev Physics Institute, 119991 Moscow, Russia \\
$^c$ Johannes Kepler Universit\"{a}t, Linz, A-4040, Austria}

%\date{\today}
\title{Interaction Effects in Conductivity of Si Inversion Layers
at Intermediate Temperatures}
\maketitle

\begin{abstract}
We compare the temperature dependence of resistivity $\rho
(T)$ of Si MOSFETs with the recent theory by Zala et al.  This
comparison does not involve any fitting parameters: the effective
mass $m^*$ and $g^*$-factor for mobile electrons have been  found
independently. An anomalous increase of $\rho$ with temperature,
which has been considered a signature of the ``metallic'' state,
can be described quantitatively by the interaction effects in the
ballistic regime. The in-plane magnetoresistance
$\rho(B_\parallel)$ is qualitatively consistent with the theory;
however, the lack of quantitative agreement indicates that the
magnetoresistance is more susceptible to the sample-specific
effects than $\rho (T)$.
\end{abstract}


\vspace{-0.1in}
\begin{multicols}{2}
The theory of quantum corrections due to single-particle (weak
localization) and interaction effects has been very successful in
describing the low-temperature electron transport in
low-dimensional conductors (see, e.g., \cite{aa}). There was,
however, a noticeable exception: the temperature- and
magnetic-field dependences of the resistivity $\rho$ of the most
ubiquitous two-dimensional (2D) system, electrons in silicon
MOSFETs, defied the theoretical predictions. Serious quantitative
discrepancies between the experimental data and the theory of
interaction corrections were noticed two decades ago (see, e.g.,
\cite{bishop,dolgopolov,burdis}). The gap between the expected and
observed low-temperature behavior of $\rho$ became dramatic with
the advent of high-mobility Si devices \cite{observation}.
Despite the earlier attempts to attribute the ``anomalous metallic behavior''
to the density- and temperature-dependent screening
\cite{screening}, the low-temperature transport
in Si MOSFETs \cite{mit}, as well as
other 2D systems with a low carrier density $n$,
resisted explanation for a decade
(for reviews, see \cite{review,review2}).

Recently, important progress has been made in both experiment
and theory, which allows to solve this long-standing problem.
Firstly, it has been recognized that multiple valleys in Si
MOSFETs enhance interaction effects \cite{punnoose,AZN}. Secondly,
the theory of interaction corrections to the conductivity
has been extended beyond the diffusive regime \cite{AZN}.
This development
is crucial because the most pronounced increase of $\rho$ with
temperature is observed in the ballistic regime (see below).
Thirdly, the Fermi-liquid interaction parameters for a dilute
electron system in Si MOSFETs have been found from detailed
measurements of Shubnikov-de Haas oscillations \cite{okamoto,gm}.
All this allows to compare the experimental data with the theory
\cite{AZN} without any adjustable parameters.

In this Letter, we show that the most prominent features of the
``metallic'' state in Si MOSFETs, the strong temperature and
in-plane magnetic field dependences $\rho(T)$ and
$\rho(B_\parallel)$ at relatively small resistivities  $\rho \ll
h/e^2$, can be accounted for by the theory \cite{AZN} over a wide
range of carrier densities, temperatures, and magnetic fields.
Thus, the ``metallic'' conductivity of high-mobility Si MOSFETs,
which was considered anomalous for a decade, can now be explained
by the interaction effects in a system with a large pseudo-spin (two
valleys + two spins), and an interaction-enhanced
spin susceptibility.
Though the theory \cite{AZN} accounts only for small corrections, it works
surprisingly well for some samples even
at relatively high
temperatures, where $\delta \rho /\rho \sim 1$.
The theory, however, does not agree with experiment at very low temperatures
and in strong magnetic fields $B_\parallel\gg T/g\mu_B$,
as well as at low electron densities, where
sample-specific effects come into play  \cite{cooldowns}.

The ac (13\,Hz) measurements of the dependences $\rho(T)$ and
$\rho(B_\parallel)$ have been performed
on six (100)~Si-MOS
samples from different wafers: Si15 (peak mobility
$\mu^{peak}=4.0$\,m$^2$/Vs), Si2Ni (3.4\,m$^2$/Vs), Si22 (3.3\,m$^2$/Vs),  Si6-14
(2.4\,m$^2$/Vs), Si43 (1.96\,m$^2$/Vs), and Si46
(0.15\,m$^2$/Vs); more detailed description of the samples can be
found in Refs.~\cite{disorder}.

The quantum corrections to the Drude conductivity
$\sigma_D=e^2n\tau/m^*$ ($\tau$ is the momentum relaxation time,
$m^*$ is the effective mass of carriers) can be expressed
(in units of $e^2/\pi \hbar$) as  \cite{AZN,aleiner}:
\begin{eqnarray}
& \sigma(T,B_\parallel)-\sigma_D = \delta \sigma_C + 15\delta\sigma_T +
 2[(\sigma(E_Z,T) - \sigma(0,T)]& \nonumber \\
& +2[\sigma(\Delta_v,T)-\sigma(0,T)]+ [\sigma(E_Z + \Delta_v,T)-\sigma(0,T)] & \nonumber \\
& + [\sigma(E_Z-\Delta_v,T)-\sigma(0,T)] + \delta\sigma_{\rm loc}(T). &
\end{eqnarray}
Here $\delta\sigma_C = x[1-\frac{3}{8}f(x)] -
\frac{1}{2\pi}\ln(\frac{E_F}{T})$ and $\delta\sigma_T  = f_1(F_0^a)
x[1-\frac{3}{8}t(x,F_0^a)]-\left(
1-f_2(F_0^a)\right)\frac{1}{2\pi}\ln(\frac{E_F}{T})$  are the
interaction contributions in the singlet and triplet channels,
respectively \cite{EF}; $\delta\sigma_{\rm loc}(T)= \frac{1}{2\pi}\ln
(\tau/\tau_\varphi(T))$ is the  weak localization contribution.
The terms $\sigma(Z,T)-\sigma(0,T)$
reduce the triplet contribution when the Zeeman energy
($Z=E_Z=2\mu_B B_\parallel$), the valley splitting
($Z=\Delta_v$), or combination of these factors ($Z= E_Z\pm
\Delta_v$) exceed the temperature.
The prefactor 15 to
$\delta\sigma_T$ reflects enhancement of the triplet contribution
due to two valleys of the electron spectrum in (100) Si MOSFETs
\cite{punnoose}. Because of this enhancement, the ``negative''
correction to the conductivity due to the triplet channel,
$d\delta\sigma_T/dT < 0$, overwhelms the ``positive'' correction
due to the singlet channel and weak localization,
$d(\delta\sigma_C+\delta\sigma_{\rm loc})/dT > 0$.
The triplet-channel prefactor might be reduced by  inter-valley
scattering when the relevant  scattering time $\tau_v<\hbar/T$.
Equation~(1)
describes the quantum corrections in both diffusive and ballistic
regimes; the crossover occurs at $T \approx (1+F_0^a)/(2\pi\tau)$
\cite{AZN}.

The terms in Eq.~(1) are functions of $x=T\tau/\hbar$, $Z$,
and $F_0^a$; their explicit expressions are given in
Ref.~\cite{AZN}. The theory \cite{AZN} considers the  Drude
resistivity $\rho_D=1/\sigma_D$ as a parameter and does not
describe the density dependence $\rho_D (n)$.
We found $\rho_D$ by extrapolation of the linear dependences $\rho(T)$
to $T=0$; $\tau$ was determined from $\rho_D$ using the
renormalized effective mass $m^*(n)$ \cite{gm}. The Fermi-liquid
parameter $F_0^a\equiv F_0^\sigma$ \cite{F0a}, which controls
renormalization of the $g^*$-factor [$g^*= g_b/(1+F_0^a)$, where
$g_b=2$ for Si], has been experimentally determined from
measurements of the Shubnikov-de Haas (SdH) effect
\cite{okamoto,gm}. Throughout the paper, we accept $\Delta_v=0$, because small
values $\Delta_v<1$K do not affect the theoretical curves at intermediate temperatures.
Thus, one can compare the experiment and the
theoretical prediction with no adjustable parameters; such comparison is the
main goal of this paper. In this important aspect, our approach differs from the attempts
to determine $F_0^a$ for $p$-type GaAs \cite{savchenko} and Si-MOSFETs \cite{vitkalov}
by fitting  $\rho(T)$ and $\rho(B_\parallel)$ with the theory \cite{AZN}.

\vspace{0.05in}
\begin{figure}
\centerline{\psfig{figure=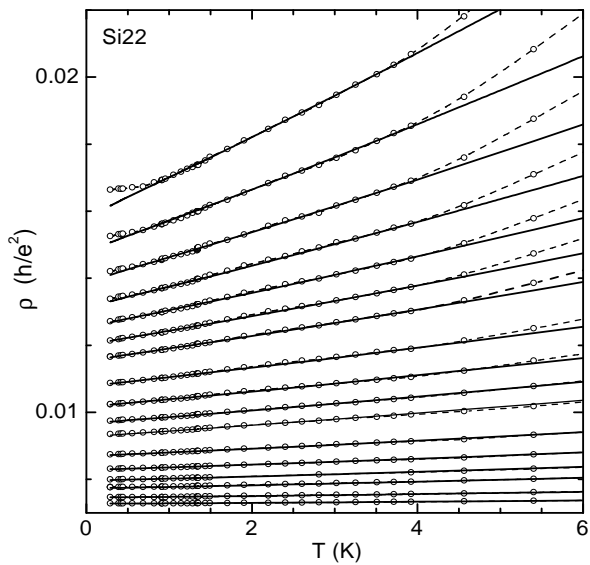,width=220pt,height=210pt}}
\vspace{0.05in}
\begin{minipage}{3.2in}
\caption{Resistivity at $B_\parallel =0$  for sample Si22 vs
temperature. The electron densities, from top to bottom are: $n=$
5.7, 6.3, 6.9, 7.5, 8.1, 8.7, 9.3, 10.5, 11.7, 12.9, 14.1, 16.5,
18.9, 21.3, 23.7, 28.5, 35.7 (in units of $10^{11}$cm$^{-2}$).
Dots and dashed lines represent the data, solid lines - the
theoretical curves with $\Delta_v=0$ and $F_0^a$ from
Ref.~\protect\cite{gm}.}
\label{fig1}
\end{minipage}
\end{figure}

Figure~1 shows the resistivity of sample Si22 versus temperature
for different electron densities. A linear dependence $\rho(T)$
extends over a decade in $T$, up to $T\approx 0.1 E_F$. In this
linear regime, the data almost coincide with the theoretical
curves (solid lines) calculated with no adjustable parameters.
Similar agreement has been observed for samples Si15 and Si43. For
the latter sample, the linear $\rho(T)$ dependences remain in
agreement with the theory up to such high temperatures ($T\sim
0.3E_F$) that $\delta\rho/\rho \sim 1$ (see Fig.~2). In this case,
which is beyond the applicability of the theoretical results
\cite{AZN}, we still calculated the corrections to the resistivity
according to $\delta\rho = -\delta\sigma \rho_D^2$ \cite{aleiner}.
For much more disordered sample Si46,  the agreement with the
theory is less impressive: the theoretical $\rho(T)$ curves are
consistent with the data only at temperatures below 10K, which
correspond for this sample mostly to the diffusive regime.

\vspace{0.05in}
\begin{figure}
\centerline{\psfig{figure=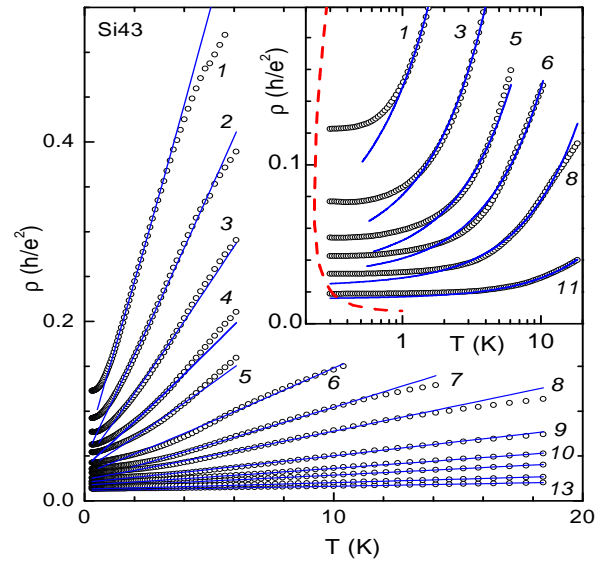,width=220pt,height=210pt}}
\vspace{0.05in}
\begin{minipage}{3.2in}
\caption{$\rho (T)$-dependences at $B_\parallel =0$ for sample
Si43 on the linear $T$-scale (main panel) and logarithmic
$T$-scale (insert).  Dots show the data, solid lines correspond to
Eq.~(1) with $\Delta_v=0$ and  $F_0^a$ from
Ref.~\protect\cite{gm}. The densities are (in units of
$10^{11}$cm$^{-2}$): {\it 1} - 1.49, {\it 2} - 1.67, {\it 3} -
1.85, {\it 4} - 2.07, {\it 5} - 2.30, {\it 6} - 2.75, {\it 7} -
3.19, {\it 8} - 3.64, {\it 9} - 4.54, {\it 10} - 5.43, {\it 11} -
6.33, {\it 12} - 8.13, {\it 13} - 9.91. The border between
diffusive and ballistic regimes is shown in the inset as the
dashed line \protect\cite{AZN}.
}
\label{fig2}
\end{minipage}
\end{figure}

For all samples,  the data agree with the theory over a broad
interval of  temperatures (see Figs.~1 and 2),
and depart from the theory on both sides of this interval. At high
temperature, the $\rho(T)$ data for Si15 and Si22 deviate from
the theory up (Fig.~1), while for Si43, they deviate mostly down
(Fig.~2). The non-universal deviations  indicate the
importance  of sample-specific  localized states  at $T\sim E_F$.
%be
%caused by thermal activation of the sample-dependent disorder.

On the low-temperature side, the dependences $\rho(T)$ tend to
saturate for all samples, in contrast to  the theoretical
prediction Eq.~(1). Weakening of the $\rho (T)$ dependence might
be caused by a non-zero valley splitting at temperatures
$T<\Delta_v$. Indeed, for samples Si22, Si15, and Si6-14,  the
saturation temperature (0.2 - 0.5\,K, see Fig.~1) is of the order
of valley splitting estimated from SdH measurements, $\sim
(0.6-0.8)$K. However, for sample Si43, the saturation temperature
is too high (1--8\,K, depending on the density - see Fig.~2),
which makes this interpretation of the saturation dubious. It is
also unlikely that the saturation at such high temperatures could
be caused by electron overheating. One of the reasons for
decrease in the interaction contribution might be a strong (and
sample-specific) inter-valley scattering. A theory which takes
inter-valley scattering into account is currently unavailable.

Finally, we note that for lower densities and  higher
resistivities $\rho \sim h/e^2$,
the slope $d\rho/dT$ changes sign. This phenomenon, known as
``metal-insulator transition in 2D''
(2D MIT), can not be accounted for by the theory \cite{AZN}; the
corresponding data for samples Si43 and Si15 across the 2D MIT
can be found in Refs.~\cite{review2,disorder}. Here we do not discuss
this non-universal and sample-dependent
regime \cite{cooldowns}.

\vspace{0.05in}
\begin{figure}
\centerline{\psfig{figure=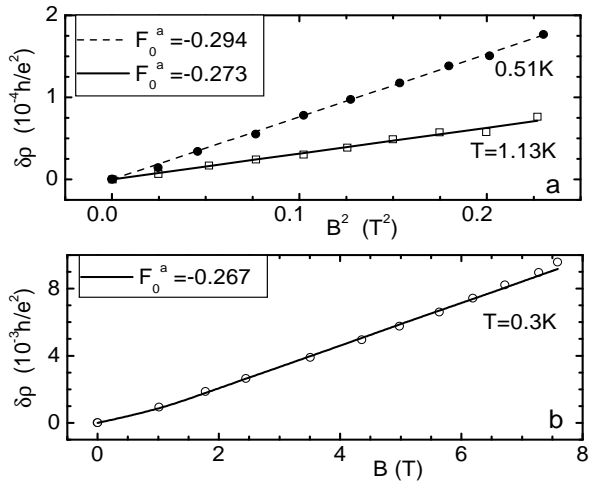,width=220pt,height=180pt}}
\vspace{0.05in}
\begin{minipage}{3.2in}
\caption{Magnetoresistivity for sample Si6-14 versus $B_\parallel^2$  at
low fields $E_Z/T <1$  (a), and versus $B_\parallel$ at high
fields $E_Z/T \gg1$ (b). The electron density $n=4.94\times
10^{11}$cm$^{-2}$ is the same for both panels. Lines are the best
fits with the $F_0^a$ values shown in the panels.}
\label{fig3}
\end{minipage}
\end{figure}

We now  turn to the magnetoresistance (MR) data. In contrast to
the temperature dependences of $\rho$, the magnetoresistance
agrees with the theory \cite{AZN} only qualitatively. For this
reason, in fitting the experimental data, we treated $F_0^a$ as
an adjustable parameter.

Figure~3 shows that the $F_0^a$ values, found for sample Si6-14
from fitting at low ($E_Z/2T <1$) and high ($E_Z/2T \gg 1$)
fields, agree with each other within 10\%; at the same time, these
values differ by ~30\% from the values determined in SdH studies
\cite{gm}. A similar situation is observed for sample Si43 in weak
and moderate fields (Fig.~4c,f and Fig.~4b,e, respectively). The
weak systematic decrease of $|F_0^a|$ with $B_\parallel$ in this
field range (Figs.~4) agrees qualitatively with non-linearity of
magnetization which we observed in SdH measurements \cite{gm}.

Fitting the weak-field MR  at different temperatures provides a
$T$-dependent $F_0^a$ (Fig.~3a). On the other hand, our SdH data
for the same sample do not confirm such a dependence: $g^*m^*$ is
constant within 2-3\% over the same temperature range \cite{gm}.
This discrepancy stems from the fact, that,
according to our data (see also Ref.~\cite{dolgopolov}),
the experimental low-field  dependence
$\rho(B_\parallel, T) \propto B^2/T^\alpha$ with $\alpha \approx 1$,
differs from the theoretical one [Eq.~(1)].

\vspace{0.05in}
\begin{figure}
\centerline{\psfig{figure=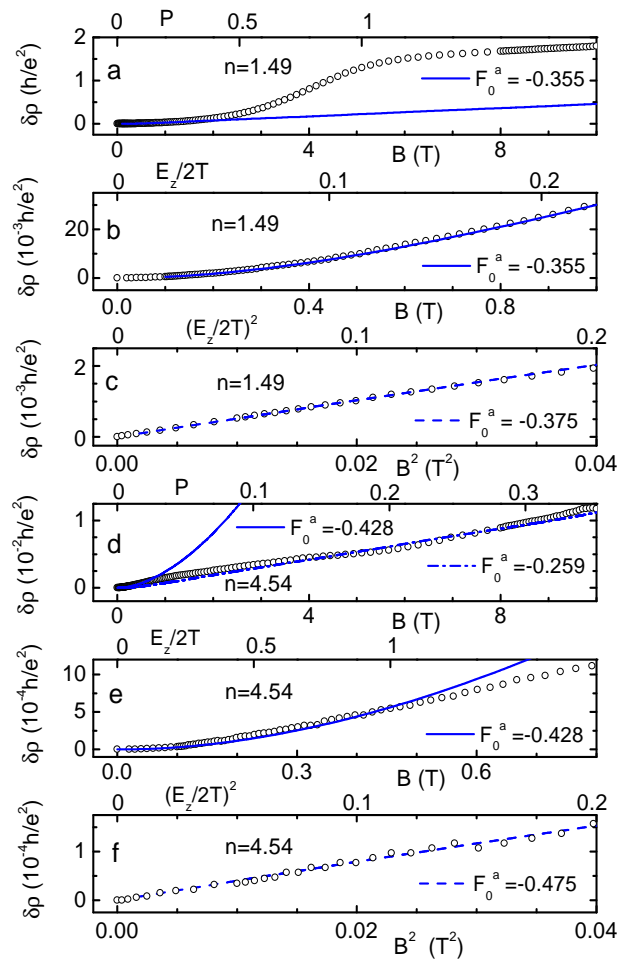,width=230pt,height=360pt}}
\vspace{0.1in}
\begin{minipage}{3.2in}
\caption{Magnetoresistivity for sample Si43 vs $B_\parallel$ and
$B_\parallel^2$ at $T=0.3$K for two densities:  $n=1.49\times
10^{11}$cm$^{-2}$ [panels a, b, c)] and $4.54\times
10^{11}$cm$^{-2}$ [panels d, e, f)]. The upper horizontal scales
 show $P\equiv g^*\mu_B B_\parallel/2E_F$ on panels $a$ and $d$, and
$E_Z/2T$ on panels
$b$, $c$, $e$, and $f$.}
\label{fig4}
\end{minipage}
\end{figure}

The discrepancy between the theory and the MR data is much more
pronounced for sample Si43 in strong fields $E_Z/2T \gg 1$, where
even the sign of deviations becomes density-dependent (compare
Figs.~4a and 4e). The non-universal behavior of the MR has been
reported earlier for different samples \cite{disorder}, and even
for the same sample cooled down to 4\,K at different fixed values
of the gate voltage \cite{cooldowns}. The reason
for this might be the interaction of mobile electrons with
field-dependent and sample-specific localized electron states.

The $F_0^a(n)$ values obtained from fitting  the low-field MR
for three samples are summarized in Fig.~5a.  The non-monotonic
density dependence of $F_0^a$ and scattering of data for different samples
indicate that the MR is more susceptible to the sample-specific
effects than $\rho(T)$.  For comparison, we present in Fig. 5\,b
the $F_0^a$-values obtained from fitting of the $\rho(T)$ data for
three samples, where we treated $F_0^a$  as a single adjustable
parameter. In contrast to Fig.~5a, there is an excellent agreement
between the $F_0^a$ values extracted from SdH measurements and
from fitting the $\rho(T)$ dependences; the agreement is observed
over a wide density range $n=(1.5 - 40)\times 10^{11}$cm$^{-2}$.

\vspace{0.05in}
\begin{figure}
\centerline{\psfig{figure=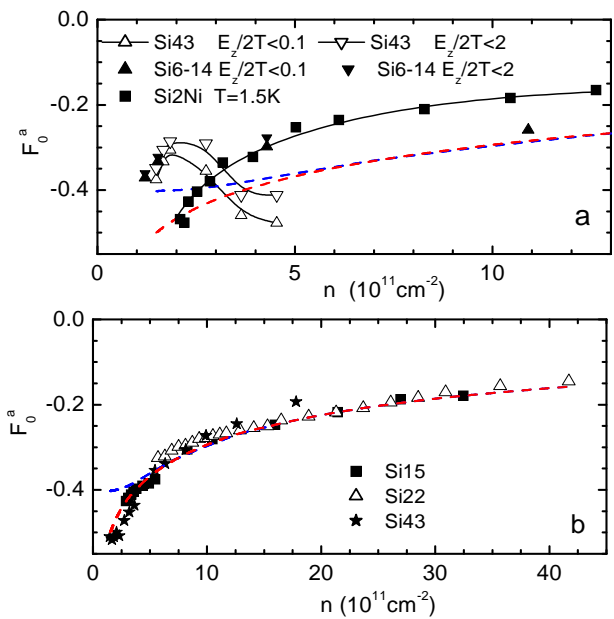,width=230pt,height=230pt}}
\vspace{0.05in}
\begin{minipage}{3.2in}
\caption{Comparison of  $F_0^a(n)$ values determined from: (a) fitting
$\rho(B_\parallel)$ for three samples, and  (b) from
fitting $\rho(T)$ for three samples. Dashed lines depict upper and
lower limits for $F_0^a$ from SdH measurements
\protect\cite{gm}.}
\label{fig5}
\end{minipage}
\end{figure}

In  summary, we performed a quantitative comparison of the
 $\rho(T,n)$ and $\rho(B_\parallel,n)$ data with the
theory which accounts for electron-electron
interactions \cite{AZN}. For high-mobility samples, we found an
excellent agreement (with no adjustable parameters) between
$\rho(T)$ and the theory in the ballistic regime over a wide range
of temperatures and electron densities $n=(1.5 - 40)\times
10^{11}$cm$^{-2}$. Our experiments strongly support the theory
attributing the anomalous ``metallic'' behavior of high-mobility Si
MOSFETs \cite{mit} to the interaction effects. Non-universal
(sample-dependent) deviations from the theory \cite{AZN} have been
observed (a) at high resistivities $\rho > 0.1h/e^2$, and (b) at
low resistivities $\rho \ll h/e^2$ for both the lowest
temperatures and high temperatures ($T\sim E_F$). The
sample-dependent deviations from the theory are more pronounced in
the in-plane magnetoresistance, especially in high fields ($2\mu_B
B_\parallel/T> 1$). We attribute this non-universality to interaction of the
mobile electrons with field-affected localized electron
states.

Authors are grateful to E.\ Abrahams, I.\ L.\ Aleiner, B.\ L.\
Altshuler, G.\ Kotliar,   D.\ L.\ Maslov, and B.\ N.\ Narozhny
for discussions. The work was supported by the NSF, ARO MURI, NWO,
NATO, FWF Austria, RFBR, INTAS, and the Russian programs ``Physics of
nanostructures'', ``Quantum and non-linear processes'',
``Integration of high education and academic research'',
``Quantum computing and telecommunications'',
and ``The State support of leading scientific
schools''.

\end{multicols}

\vspace{-0.2in}
\begin{references}
\vspace{-0.5in}


\bibitem{aa} B.\ L.\ Altshuler, A.\ G.\ Aronov, P.\ Lee Phys. Rev. Lett. {\bf 44}, 1288  (1980).

\bibitem{bishop} D.\ J.\ Bishop, R.\ C.\ Dynes, and D.\ C.\ Tsui, Phys. Rev.
B {\bf 26}, 773 (1982).

\bibitem{dolgopolov} V.\ T.\ Dolgopolov, S.\ I.\ Dorozhkin, and
A.\ A.\ Shashkin, Solid State Commun. {\bf 50}, 273 (1984).

\bibitem{burdis} M.\ S.\ Burdis and C.\ C.\ Dean, Phys.
Rev. B {\bf 38}, 3269 (1988).

\bibitem{observation}M.\ D'Iorio, V.\ M.\ Pudalov, and S.\ G.\ Semenchinsky,
Phys. Lett A {\bf 150}, 422 (1990).


\bibitem{screening} F.\ Stern, Phys. Rev. Lett. {\bf 44}, 1469 (1980).
F.\ Stern, S.\ Das Sarma, Solid State Electron. {\bf 28}, 158 (1985).
A.\ Gold, V.\ T.\ Dolgopolov, Phys. Rev. B {\bf 33}, 1076 (1986).
S.\ Das Sarma, E.\ H.\ Hwang, Phys. Rev. Lett. {\bf 83}, 164 (1999).

\bibitem{mit}S.\ V.\ Kravchenko, G.\ V.\
Kravchenko,  J.\ E.\ Furneaux, V.\ M.\ Pudalov, and M.\ D'Iorio,
\prb {\bf 50}, 8039 (1994).


\bibitem{review} E.\ Abrahams, S.\ V.\ Kravchenko, and M.\ P.\
Sarachik, Rev. Mod. Phys. {\bf 73}, 251 (2001).

\bibitem{review2}B.\ L.\ Altshuler, D.\ L.\ Maslov, and V.\ M.\ Pudalov
Physica E, {\bf 9}(2) 209-225 (2001).

\bibitem{punnoose}A.\ Punnoose, A.\ M.\ Finkelstein, Phys. Rev.
Lett. {\bf 88}, 016802 (2002).

\bibitem{AZN}G.\ Zala, B.\ N.\ Narozhny, and I.\ L.\ Aleiner. Phys. Rev. B
{\bf 64}, 214204 (2001);
Phys. Rev. B {\bf 65}, 020201 (2001).

\bibitem{okamoto}T.\ Okamoto, K.\ Hosoya, S.\ Kawaji, and A.\ Yagi,
Phys. Rev. Lett. {\bf 82}, 3875 (1999).

\bibitem{gm} V.\ M.\ Pudalov, M.\ Gershenson, H.\ Kojima, N.\ Butch, E.\ M.\
Dizhur, G.\ Brunthaler, A.\ Prinz, and G.\ Bauer, Phys. Rev.
Lett. {\bf 88}, 196404 (2002). The upper and lower limits
for the $g^*$-factor over the range $r_s=1.6 - 8.3$ are:\\
$g^*_{\rm high} = 2 + 0.3485r_s -0.01068r_s^2 +0.00048r_s^3$ and \\
$g^*_{\rm low}=2 + 0.1694r_s + 0.1233r_s^2 - 0.03107r_s^3+0.002r_s^4$.

\bibitem{cooldowns}V.\ M.\ Pudalov, M.\ E.\ Gershenson, H.\ Kojima,
cond-mat/0201001.

\bibitem{disorder} \ M.\ Pudalov, G.\ Brunthaler,
A.\ Prinz, G.\ Bauer, cond-mat/0103087; Phys. Rev. Lett. {\bf
88}, 076401 (2002).

\bibitem{aleiner} I.\ L.\ Aleiner, B.\ N.\ Narozhny, private communication.

\bibitem{EF} Throughout the paper, we use the unrenormalized Fermi energy
$E_F$.

\bibitem{F0a}$F_0^a = -F$, in notations of  Refs.~\protect\cite{aa,bishop}.
\bibitem{savchenko}Y.\ Y.\ Proskuryakov, A.\ K.\ Savchenko, S.\ S.\ Safonov,
M.\ Pepper, M.\ Y.\ Simmons, and D.\ A. Ritchie, cond-mat/0109261.

\bibitem{vitkalov} A.\ A.\ Shashkin, S.\ V.\ Kravchenko, V.\ T.\ Dolgopolov, T.\ M.\ Klapwijk,
cond-mat/0111478. 
S.\ A.\ Vitkalov, K.\ James, B.\ N.\ Narozhny, M.\ P.\ Sarachik,
T.\ M.\ Klapwijk, cond-mat/0204566.


\end{references}
\end{document}